# Morbidity, mortality and the illness-death model

Ralph Brinks

Institute for Biometry and Epidemiology, German Diabetes Center, and

Hiller Research Unit for Rheumatology, University Hospital

Düsseldorf, Germany

## Introduction

In many countries of the world, the life expectancy (LE) at birth is increasing. For instance, in Germany, LE at birth has more than doubled in the past 140 years. The question arises if the decrease in mortality is accompanied with decreased amount of life time with morbidity. James Fries introduced the idea that longer length of life comes along with a shortening of the period of morbid life, which he termed "compression of morbidity" (COM) [Fries 1980]. In a later keynote, Fries distinguishes between absolute and relative COM. Absolute COM refers to a shorting of the life expectancy with morbidity (LEM) whereas relative COM refers to a decreasing quotient LEM over LE [Fries 1982]. Until now the hypothesis of COM has inspired many researchers to study health trends.

Obviously, the questions about the COM hypothesis refer to three different health-related states: A subject can either be *Healthy*, *Morbid* or *Dead*. During the course of life, people may change the states according the transitions shown in Figure 1. The incidence rate $\lambda$ of morbidity and the mortality rates $\mu_0$ and $\mu_1$ depend on two different time scales, calendar time *t* and age *a*.

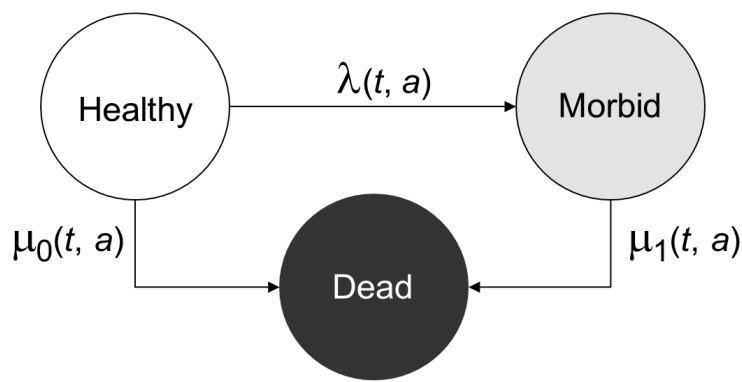

Figure 1: Illness-death model. Subjects in the population can either be *Healthy*, *Morbid* or *Dead*. The transition rates between the states (incidence $\lambda$ and the mortality rates $\mu_0$, $\mu_1$) depend on the calendar time *t* (period) and on the age *a*.

Recently, it has been shown that the proportion (prevalence *p*) of people in the *Morbid* state aged *a* at time *t* can be described by a partial differential equation [Brinks 2014, Brinks 2015]. These recent results are used in this article to present a rigorous mathematical formulation of the COM in the framework of the illness-death model.

## Analytical considerations

Life expectancy at birth of a birth cohort born at $t_0$, can be expressed by the (overall) mortality µ (Equation 1).

$$\text{LE}(t_0) = \int_0^\infty \exp\left(-\int_0^\tau \mu(t_0 + x, x)\mathrm{d}x\right)\mathrm{d}\tau \tag{1}$$

Using the prevalence *p* of morbidity, the overall mortality µ can be expressed as $\mu = p \times \mu_1 + (1-p) \times \mu_0$ which yields Equation (2).

$$\text{LE}(t_0) = \int_0^\infty \exp\left(-\int_0^\tau \{p \cdot \mu_1 + (1-p) \cdot \mu_0\}(t_0 + x, x)\mathrm{d}x\right)\mathrm{d}\tau \tag{2}$$

The LEM can be expressed in a similar way (Equation 3), which follows from Sullivan's Equation for the Healthy Life Expectancy (HLE) [Sullivan 1971] and the fact that LE equals the sum of HLE and LEM, LE = LEM + HLE.

$$\text{LEM}(t_0) = \int_0^\infty p(t_0 + \tau, \tau) \cdot \exp\left(-\int_0^\tau \{p \cdot \mu_1 + (1-p) \cdot \mu_0\}(t_0 + x, x)\mathrm{d}x\right)\mathrm{d}\tau \tag{3}$$

In [Brinks 2014, Brinks 2015] it has been proven, that the prevalence *p* of morbidity is linked to the rates in the IDM (Figure 1) by the partial differential equation (PDE) as shown in Equation (4)

$$\left(\frac{\partial}{\partial t} + \frac{\partial}{\partial a}\right)p = (1-p)\{\lambda - p(\mu_1 - \mu_0)\} \tag{4}$$

Usually, the PDE (4) is combined with an initial condition which guarantees the existence of a unique solution if certain smoothness constraints of the rates $\lambda$, $\mu_0$, and $\mu_1$ are fulfilled [Polyanin 2001]. Henceforth, we will use the initial condition that morbidity is contracted after birth, i.e. $p(t, 0) = 0$ for all *t*. Since there are congenital anomalies with severe morbidity, this assumption is only an approximation to reality. However, most congenital diseases are relatively rare.

Since LE and LEM (Eq. (2) and (3)) only depend on the prevalence *p* and the mortality rates $\mu_0$ and $\mu_1$ and *p* in turn depends on the rates $\lambda$, $\mu_0$, and $\mu_1$ from the IDM via Equation (4), we see that the rates $\lambda$, $\mu_0$, and $\mu_1$ completely determine the temporal trend in LE and LEM. With other words, the mathematical properties of the rates from IDM determine if the hypothesis of COM is fulfilled or not.

# Example

To demonstrate the usefulness and feasibility of the theoretical framework of the previous section, we present a hypothetical example motivated from Germany. We start from the projected mortality rate µ for men from the German Federal Statistical Office. Based on published values of µ from 2012 to 2060 [Federal Statistical Office 2009], we fit a linear regression $\log(\mu)(t, a) = \beta_{00} + \beta_{01} a + \beta_{10} t + \beta_{11} t a + \beta_{02} a^2$ which gives a reasonably good fit to the empirical values. Calendar time *t* is measured in years after 2000, age *a* is measured in years. The mortality $\mu_0$ is assumed to be $\mu_0(t, a) = \gamma_0 \{\exp(\gamma_1 (a - 65)_+) - 1\}$ where $x_+$ means the positive part of x, i.e., $x_+ = \max(0, x)$. The choice of the $\mu_0$ is motivated by the description of the "ideal survivorship curve" (Figure 5 in [Fries 1982]).

The incidence λ is assumed to be $\lambda(t, a) = \exp(\delta_0 + \delta_1 a + t \log(1+\delta_2))$.

In the situation that λ, µ and $\mu_0$ are given, the prevalence *p* from (4) with initial condition $p(t, 0) = 0$ has a closed form representation as given in Equation (5).

$$p(t,a) = 1 - \exp\left(-\int_0^a \{\lambda - \mu + \mu_0\}(t - a + \tau, \tau) \mathrm{d}\tau\right) \qquad (5)$$

This can be derived by using that the right rand side of the PDE (4) equals $(1 - p)(\lambda - \mu + \mu_0)$. Then, Eq. (4) becomes linear.

The integral in Eq. (5) can be calculated by Romberg's Method [Dahlquist 1974]. The resulting age-specific prevalences in the age range 50 to 100 years in the years *t* = 0 (blue) and *t* = 10 (black) are presented in Figure 2. We see that in the higher age groups there is a slight increase from *t* = 0 to *t* = 10.

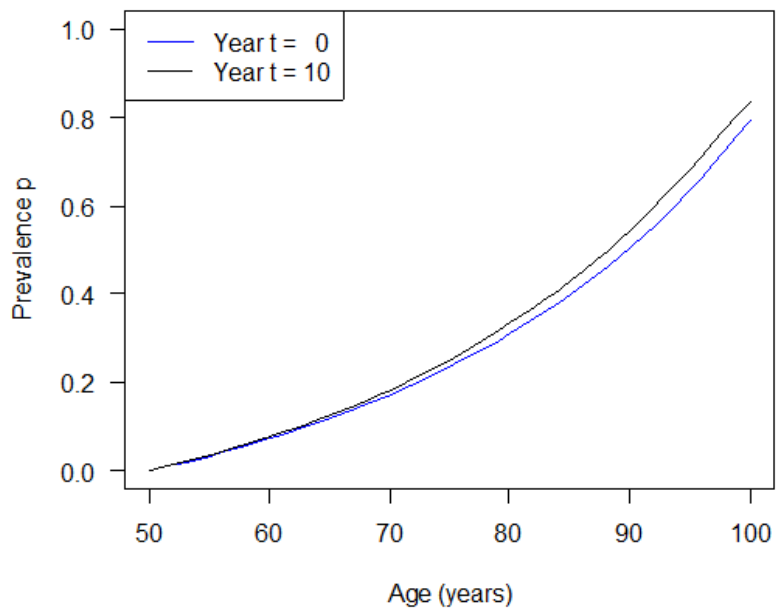

**Figure 2: Age-specific prevalences *p* for the years *t* = 0 (blue line) and *t* = 10 (black line).**

Inserting *p* into (2) and (3), yields the values presented in Table 1. Note that we refer to the age 50. We see that the LE at age 50 in year 0 is 31.5 years. Ten years later (year 10), LE has increased by 1.3 years to 32.8 years. LEM at age 50 has also increased from 5.40 to 6.29 years during this period. Thus, we do not have an absolute compression of morbidity. From the last row in Table 1 we can also see that we do not have a relative COM either. In fact, the percentage of life spent in morbidity increases from 17.1% to 19.1% during the considered time period.

Table 1: Life expectancy (LE), life expectancy with morbidity (LEM) and proportion of morbid life (LEM/LE) for the years $t_0 = 0$ and $t_0 = 10$ in the example described in the text.

|  | Year | |
|---|---|---|
|  | $t_0 = 0$ | $t_0 = 10$ |
| Life expectancy (LE) (years) at age 50 | 31.5 | 32.8 |
| Life expectancy with morbidity (LEM) at age 50 | 5.40 | 6.29 |
| LEM/LE | 17.1% | 19.1% |

# Conclusion and discussion

In this article, we use the illness-death model to present a mathematical framework for studying the compression of morbidity (COM) hypothesis. It turns out, that the COM is closely related to the well-known illness-death model, which allows a rigorous analytical examination. While the link between the COM and the IDM is not new [Varadhan 2014], this is the first time that the COM hypothesis is expressed in mathematical equations. Thus, the COM hypothesis has been made tractable by analytical tools, such as advanced calculus and numerical analysis.

To demonstrate the usefulness of the mathematical framework, an example was given, which has been motivated by empirical findings from Germany. In the example, the life expectancy (LE) increased as well as the life expectancy with morbidity (LEM). However, the relative gain in LEM exceeded the relative gain in LE, which implies that the proportion of morbid life (LEM/LE) increases. In the terminology of Fries, in the example there is neither an absolute nor a relative COM.